\documentclass{emulateapj}
\usepackage{color}
\usepackage{times}

\shortauthors{GREEN \& KLIEM}
\shorttitle{FLUX ROPE FORMATION PRECEDING CME ONSET}  

\begin{document}

\title{Flux Rope Formation Preceding Coronal Mass Ejection Onset}

\author{L. M. Green \altaffilmark{1}
        and
	B. Kliem\altaffilmark{1,2,3}}
\affil{$^1$University College London, Mullard Space Science Laboratory,
           Holmbury St.\ Mary, Dorking, Surrey, RH5 6NT, UK}
\affil{$^2$Naval Research Laboratory, Space Science Division,
           Washington, DC 20375, USA}
\affil{$^3$University of Potsdam, Institute of Physics and Astronomy,
           Potsdam 14476, Germany}

\email{lmg[bhk]@mssl.ucl.ac.uk}

\journalinfo{ApJL, in press}
\submitted{Received 2009 April 27, accepted 2009 June 22}

\begin{abstract}

We analyse the evolution of a sigmoidal (S shaped) active region toward
eruption, which includes a coronal mass ejection (CME) but leaves part of
the filament in place. The X-ray sigmoid is found to trace out three
different magnetic topologies in succession: a highly sheared arcade of
coronal loops in its long-lived phase, a bald-patch separatrix surface
(BPSS) in the hours before the CME, and the first flare loops in its
major transient intensity enhancement. The coronal evolution is driven by
photospheric changes which involve the convergence and cancellation of
flux elements under the sigmoid and filament. The data yield unambiguous
evidence for the existence of a BPSS, and hence a flux rope, in the
corona prior to the onset of the CME.

\end{abstract}


\section{INTRODUCTION}
\label{sec:Introduction}

Coronal mass ejections (CMEs) are large scale eruptions of magnetized
plasma from the solar atmosphere into the interplanetary space. It is
generally accepted that their energy source is derived from the free
energy contained in sheared or twisted magnetic fields
\citep{Forbes2000}. Many CME models have been developed and despite the
differences in the underlying physics of the eruption, all models at some
point involve a magnetic flux rope. From this viewpoint, the models can
be split into those which require the flux rope to exist prior to the
eruption, and those in which the flux rope is formed as a result of
topological changes in the course of the eruption.

In the first category the rope is fundamental to the CME initiation
process. These models include ideal MHD instabilities of a flux rope
\citep{Torok&Kliem2005, Kliem&Torok2006}, the force imbalance between a
rope and its overlying arcade-like field
\citep{Mackay&vanBallegooijen2006, Bobra&al2008}, and the catastrophe of
the rope-arcade configuration as a whole \citep{Forbes&Isenberg1991}. In
the second category, it is proposed that a sheared magnetic arcade
becomes unstable to upward expansion \citep{Antiochos&al1999,
Moore&al2001}. In the course of the expansion, a vertical current sheet
forms in the bottom part of the arcade. Magnetic reconnection in the
sheet transforms the inner part of the arcade into a growing flux rope
starting early in the expansion process \citep{Lynch&al2008} and
supporting the expansion in a positive feedback. Obviously, determining
whether the pre-eruption magnetic field topology involves a flux rope is
key to understanding the physics of CME initiation.

How a flux rope can be formed in the corona prior to an eruption is an
open question as well. It may emerge from the convection zone
\citep{Rust&Kumar1994, Low1996} or be formed in situ by an arcade-to-rope
topology transformation \citep{vanBallegooijen&Martens1989}. Numerical
simulations of flux emergence indicate that the process essentially stops
as the magnetic axis of the rope hits the photosphere, producing a
sheared arcade in the corona \citep{Fan2001}. Moreover, pre-CME formation
of large-scale flux ropes between active regions and in the quiet Sun
cannot occur by emergence. The theoretical predictions for the evolution
of an emerged arcade range from further continuous shearing, keeping the
arcade topology until an eruption occurs after 1--2 weeks
\citep{vanBallegooijen&Mackay2007}, to the immediate formation of a
coronal flux rope, which can find a stable equilibrium or erupt readily,
depending on the relative strength and orientation of emerging and
preexisting field \citep{Manchester&al2004, Archontis&Torok2008}. The
observations indicate that active regions typically produce eruptions
several days after their emergence is largely complete. This is
consistent with both, gradual arcade shearing and gradual arcade-to-rope
topology transformation \cite[e.g.,][]{Amari&al2003a, Amari&al2003b}.
Dispersal and diffusion of the photospheric flux concentrations and flows
converging toward the polarity inversion line (PIL) of the photospheric
field are main drivers of such quasi-static coronal evolution, which
often involves flux cancellation in the photosphere along the PIL.

The observational evidence for the existence of a flux rope topology is
most compelling for the evolved stage of a CME and gets weaker as earlier
phases are considered. Interplanetary data reveal that at 1~AU many CMEs
are well described by a flux rope model \citep{Jian&al2006} and coronal
images often suggest a flux rope topology for erupting filaments and CME
cores. Some erupting filaments exhibit a rotation of their axis about the
direction of ascent
\citep{Rust&LaBonte2005, Green&al2007};
this results from a
conversion of twist to writhe and implies a flux rope topology. On the
other hand, it is unclear whether filaments in their stable equilibrium
prior to eruption are contained in fields of flux rope
\citep{Rust&Kumar1994} or arcade topology \citep{Martin1998} as it is not
yet possible to directly measure the three-dimensional structure of the
coronal field. Measurements at the photospheric boundary have supported
the flux rope topology, however \citep{Lites2005}.

The pre-eruption flux rope formation processes mentioned above involve
reconnection at or above the PIL, which, in the case of gradual evolution,
is presumably weak and
intermittent and in principle not different from the ubiquitous
small-scale reconnection events in the solar atmosphere. Therefore, to
our knowledge, it has not yet been possible to infer the formation of a
flux rope from signatures of reconnection during the quasi-static
evolution prior to eruptions. Support was obtained from the observation
of an S shaped emission source at low coronal temperatures, interpreted
as cooling plasma on newly reconnected field lines within a flux rope
\citep{Tripathi&al2009}. Events of flux
convergence and cancellation at a large-scale PIL presumably represent
the clearest indication of the formation process, but are no proof.
Another indication, though not a proof either, is the ``necking'' of
coronal cavities in the final stages of their evolution toward eruption
\citep{Gibson&al2006a}. Here we present an investigation of coronal
reconnection signatures that demonstrate the formation of a flux rope
prior to the onset of a CME.

\section{BALD-PATCH SEPARATRIX SURFACE}
\label{sec:BPSS}

We study a sigmoidal active region (S shaped in X-rays) from the
quasi-static evolution, appearing as a long-lived sigmoid, to its
eruption, which produced two topologically different stages of a
transient sigmoid. Such sources are a signature of locally enhanced
dissipation---most likely reconnection, and their occurrence is 
correlated with CMEs \citep{Canfield&al1999}. Transient sigmoids are
formed in quasi-separatrix layers (QSLs), which represent the interface
between two topologically distinct flux systems
\citep{Priest&Demoulin1995}; in the case of sigmoids a flux rope and its
surrounding arcade \citep{Titov&Demoulin1999, Green&al2007}. This may
also be true for long-lived sigmoids \citep{Gibson&al2006b};
alternatively, these may just represent a collection of highly sheared
coronal loops, i.e., an arcade \citep{Moore&al2001}.

The QSL in a flux rope-arcade system can be of two different topologies.
It can either intersect itself under the rope along a line of magnetic
X-type topology, or extend down to the bottom of the atmosphere
\citep{Titov&Demoulin1999}.
The latter situation is
expected for the emergence of a sub-photospheric flux rope's full cross
section as suggested in \cite{Low1996}. Which of the two topologies
results from the process of arcade-to-rope transformation, depends upon
how high in the atmosphere the reconnection proceeds and whether
all reconnected flux under the X line submerges below the photosphere,
dragging the X line downward. Both are open questions. We will focus on
the topology without an X line, since this is relevant to the data
investigated here.

In the absence of an X-type topology under the rope, the QSL touches the
PIL tangent to the photosphere, with its field lines pointing in the
inverse direction (from the negative to the positive side of the PIL).
Sections of the PIL having this property appear as bald patches in
H$\alpha$ images, so the corresponding QSL has become known as bald-patch
separatrix surface, BPSS \citep{Gibson&Fan2006}. The field lines passing
through bald patches must wrap around the flux rope to connect to
their end points on the other side of the PIL
\cite[see, e.g.,][Fig.~3]{Gibson&al2006b}. Hence, they are
sigmoidal in shape, being forward (reverse) S for right-handed
(left-handed) chirality.

In the process of flux rope formation by arcade-to-rope transformation or
by rope emergence, the bald-patch field lines have the highest likelihood
of all the field lines in the configuration to be illuminated in X-rays
because they pass through the volume of reconnection, at least
temporarily. The hot plasma
resulting from reconnection expands in the corona along these field
lines.

Due to inertial line tying in the photosphere, the bottom of the BPSS
must stay largely immobile at coronal timescales. However, the top part
of the flux rope can tear apart from the bottom section. This makes the
BPSS topology relevant for so-called partial eruptions which produce a
CME but leave behind filament material in the bottom part of the source
volume \citep{Gibson&Fan2006}. In the event of eruption, the bottom part
of the BPSS survives and is likely to be illuminated further as the
rising upper part of the flux rope tears off, causing a steepening of the
currents in the BPSS and, correspondingly, further dissipation. However,
a partial flux rope eruption also forms a vertical current sheet above
the PIL and inside the original rope, i.e., topologically distinct from
the BPSS. Once formed, it becomes the place of strongest energy release,
and the flare loops emerging in its downward reconnection outflow are
likely to become the brightest X-ray source.

From the above it is clear that a sigmoid can be expected to occur as a
signature of flux rope formation in the corona and, reversely, that the
observation of a BPSS sigmoid implies a flux rope topology. This sigmoid
must be a continuous S crossing the PIL three times, different from a
sheared arcade whose loops generally cross the PIL only once. When the
flux rope experiences a partial eruption, the middle part of the BPSS
sigmoid remains stationary, while a new, brighter X-ray source appears
above it, which evolves into the post-eruption arcade.

\section{SIGMOID AND PHOTOSPHERIC FIELD}
\label{sec:Observations} 

\begin{figure}[t]                                                
\centering
\includegraphics[width=1.0\columnwidth]{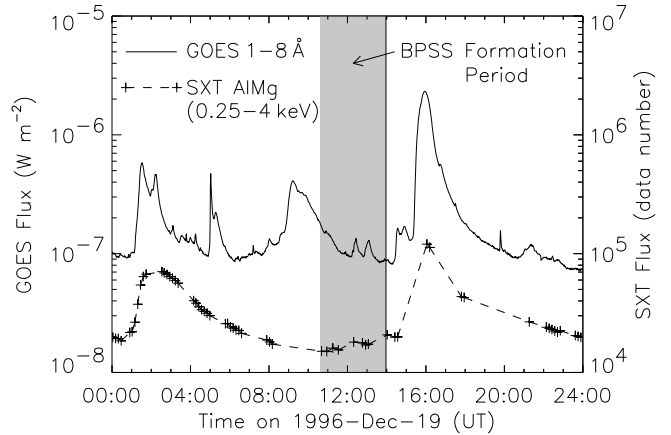}
\caption{
 Light curves of the flares at 01:22 and 15:21~UT
 (from \textsl{GOES} and constructed from SXT/AlMg images).
 The precursors after 14:27~UT also occurred in AR~8005, whereas the
 flares at 04:57 and 08:23~UT had different locations.
\label{fig:lightcurves}}
\end{figure}

NOAA active region (AR) 8005 hosted a foward S sigmoid and produced a CME
on 1996 December 19 which was associated with a \textsl{GOES} C2.3 class
X-ray flare at 15:21~UT (Figure~\ref{fig:lightcurves}). We study the
coronal and photospheric evolution leading to the event using,
respectively, half and quarter resolution AlMg and Al.1 filter images
recorded by the Soft X-ray Telescope (SXT) onboard \textsl{Yohkoh}
\citep{Tsuneta&al1991} and magnetograms obtained by the Michelson Doppler
Imager (MDI) onboard \textsl{SOHO} \citep{Scherrer&al1995}.

\begin{figure*}[t]                                                
\centering
\includegraphics[width=.95\textwidth]{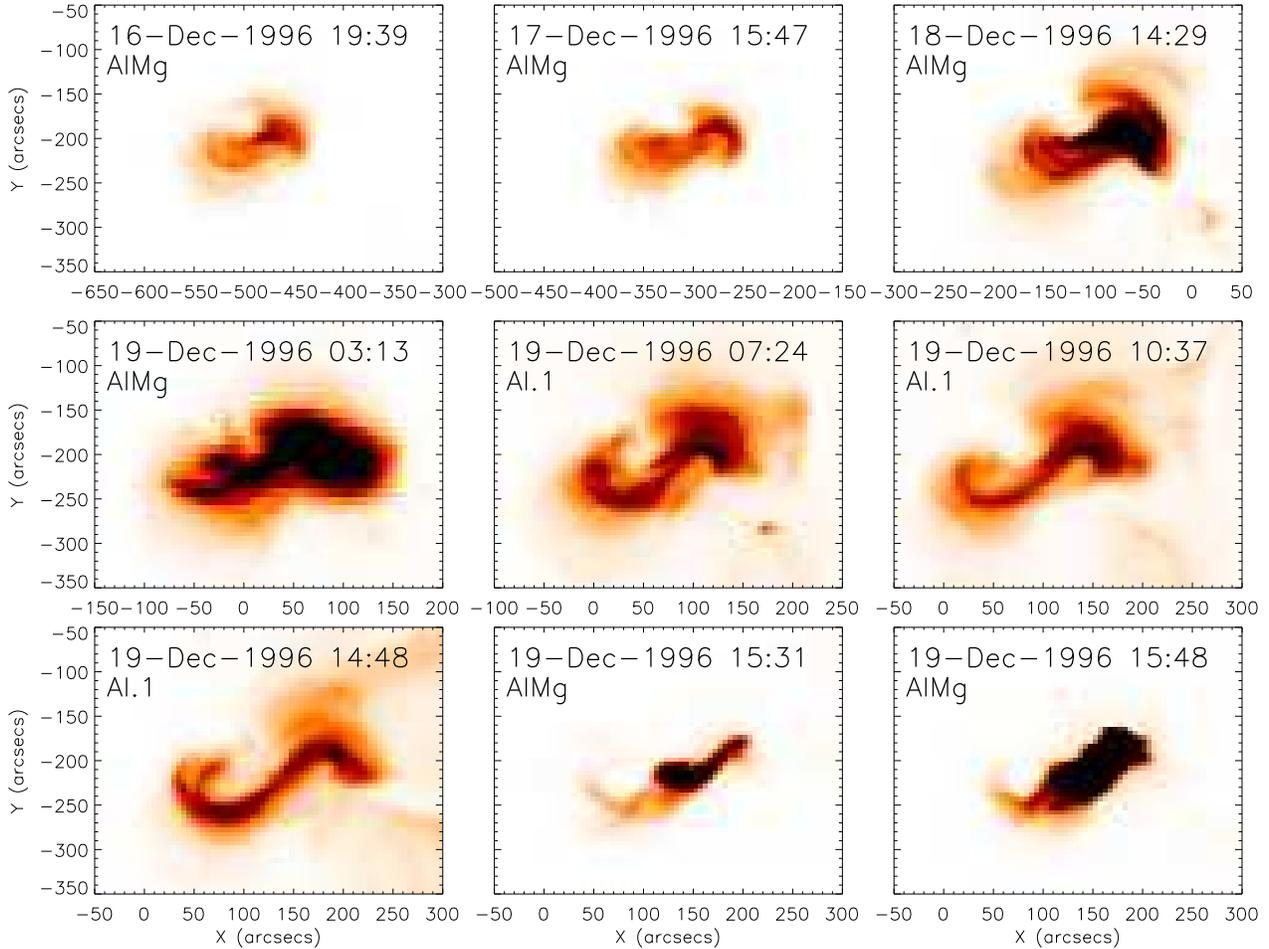}
\caption{
 SXT images showing the evolution of AR~8005 from a sheared arcade, to
 point symmetric J shaped loops, to a continuous sigmoid structure, and
 finally into a sigmoidal flare emission core that grows into the
 post-eruption arcade.
\label{fig:sigmoid_evolution}}
\end{figure*}

The active region was in its decay phase as it rotated over the limb,
already having dispersed magnetic polarities and no sunspots. In the
three days leading up to the event, the X-ray loops showed a gradual
evolution to larger extent, whilst the appearance of a sheared arcade
with an overall S shape was maintained
(Figure~\ref{fig:sigmoid_evolution}). This appearance results from the J
shape of the individual loops, produced as a direct consequence of the
accumulation of longitudinal flux (pointing along the PIL) in the centre
of active regions in the course of their long-term, largely diffusive
evolution \citep{vanBallegooijen&Mackay2007}. The dominance of this flux
gives all arcade field lines in the middle of the region a straight leg
and permits them to pass over the PIL only at the edges of the region,
producing the J shape.

Beginning at about 00~UT on December~19, the sigmoid narrowed. At
01:22~UT a weak, non-eruptive flare occurred near its northwestern end.
After the flare had faded, the sigmoid consisted of two sets of sharper
J shaped loops (threads) with shortened straight legs. By 10:37~UT the
first J shaped threads had merged into a continuous S, and by the
exposure at 13:57~UT the evolution of all visible threads into a
continuous S shaped sigmoid was complete
(Figure~\ref{fig:sigmoid_evolution}). As justified below, we regard this
to be a BPSS sigmoid. It had a dip in intensity in its central portion.

The SXT images taken during the main rise of the flare (15:31--15:48~UT)
show that the continuous S shaped sigmoid survived the onset of the
eruption. Its eastern elbow remained visible under the growing arcade of
flare loops for at least two hours, well into the flare decay phase.
Moreover, the sigmoid showed no signatures of motion. Its elbows turned
around less and shrank somewhat in the course of the flare, but the
central part remained stationary. The rise of the soft X-ray flux
originated in a new source located in the core of the preexisting
sigmoid, overlying its dip. The new source was closely aligned to the
preexisting sigmoid's central section, but with strands seen branching
off, so initially it formed a new sigmoid with clearly different end
points. It evolved into the post-eruption arcade, which suggests that the
new sigmoid was composed of the first, highly sheared flare loops, formed
in the downward reconnection outflow of a vertical current sheet. The
evolution of the new sigmoid into the post-eruption arcade was described
in greater detail by \cite{Sterling&al2000}, who also concluded
that it was a structure different from the preexisting sigmoid.

H$\alpha$ and Helium 10830~{\AA}
data\footnote{\texttt{ftp://nsokp.nso.edu};
\texttt{http://www.bbso.njit.edu}}
reveal the presence of a filament situated along the PIL, whose  central
and southeastern section survived the eruption, as seen in the available
images at 15:52, 16:57, and 20:33~UT.

\begin{figure*}[t]                                                
\centering
\includegraphics[width=.95\textwidth]{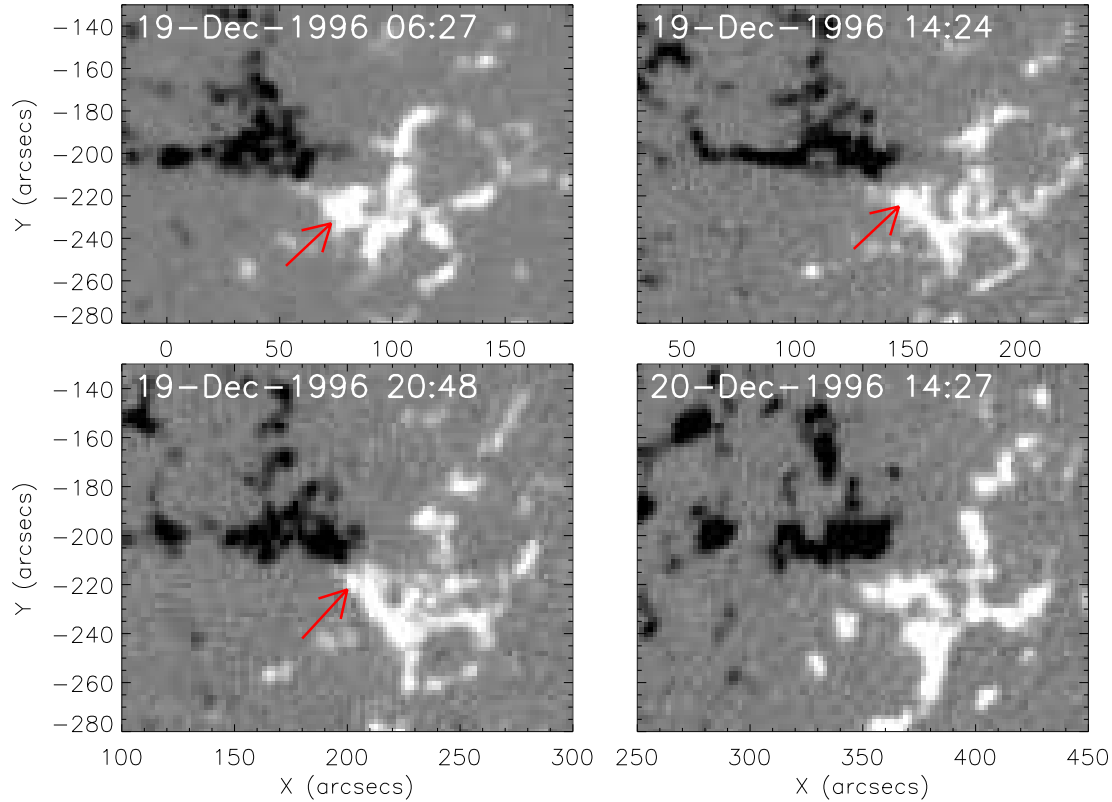}
\caption{
 MDI magnetogram sequence (saturated at $\pm$300~Gauss) showing the flux
 cancellation episode associated with the formation of the continuous
 sigmoid. A region of strong positive flux extends and moves toward the
 PIL (indicated by arrows). When the polarities meet, cancellation occurs
 and produces a channel of low flux density.
  An animation of the MDI data throughout December 15--20 is available in
  the electronic edition of the Journal.
\label{fig:cancellation}}
\end{figure*}

The evolution of the photospheric flux was dominated by ongoing dispersal
of the main flux concentrations, with no significant emergence of new
flux, throughout the disk passage of the active region (see the
animated MDI data accompanying Figure~\ref{fig:cancellation}).
This included several episodes in
which patches of strong flux detached from the main polarities,
approached the PIL, and cancelled. One of these occurred at the
northwestern end of the sigmoid between about December~18, 21~UT and
December~19, 08~UT and was obviously associated with the small preceding
flare. Another, stronger cancellation episode occurred between about
December~19, 03~UT and December~20, 11~UT in a narrow region under the
middle of the sigmoid, where the two J's merged into the continuous S
(Figure~\ref{fig:cancellation}). It was preceded by cancellations of weak
flux elements in the same area since about December~18, 20~UT. These data
indicate that the changes in the sigmoid were not causally related to the
preceding flare, but were driven primarily by the cancellation episode
shown in Figure~\ref{fig:cancellation}.

\begin{figure*}[t]                                                
\centering
\includegraphics[width=1.0\textwidth]{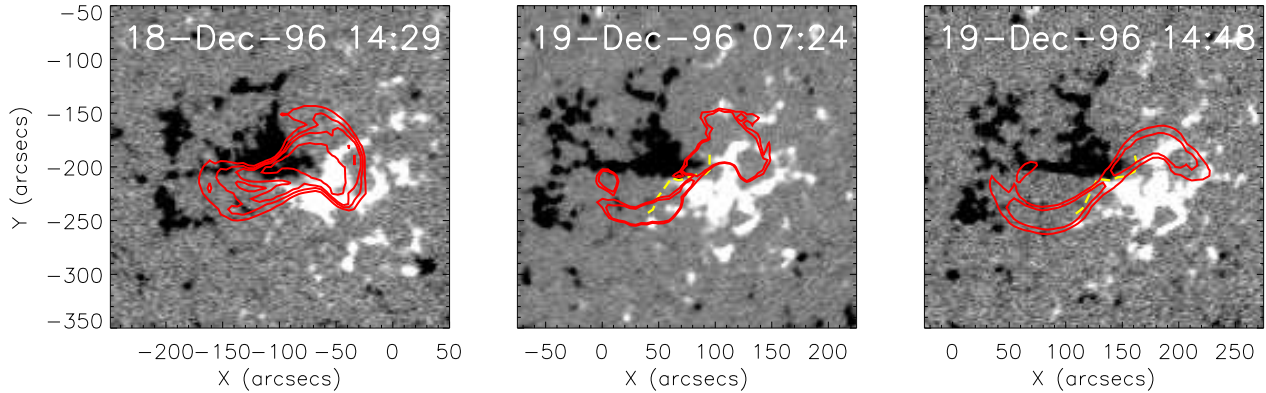}
\caption{
 Co-aligned MDI and SXT data showing the magnetic connections of
 the sheared arcade, double-J loops, and continuous sigmoid,
 and the continuous sigmoid's triple crossing of the PIL (dashed yellow
 line). The nearest magnetogram in time (saturated at $\pm$100~Gauss)
 was differentially rotated to the time of the SXT data. 
\label{fig:mdi_overlays}}
\end{figure*}

Co-aligned MDI and SXT data on December 18--19 show that each of the two
sets of point symmetric J shaped loops made one crossing of the PIL and
had clearly discernible end points at either side of it.
When the continuous sigmoid formed,
it made three PIL crossings; twice by the sigmoid elbows and once as its
central part crossed the PIL in the inverse direction
(Figure~\ref{fig:mdi_overlays}). Overall, the PIL ran at about
$-45^\circ$ inclination to the meridional lines, parallel to the central
part of the sigmoid. This is a well known characteristic of sigmoids
which so far has prevented the proof of an inverse PIL crossing. However,
at the point where the major flux cancellation on December 19--20
occurred and the continuous sigmoid formed, the PIL ran locally nearly
east-west, at a substantial inclination to the sigmoid at that point.

\section{IMPLICATIONS FOR FLUX ROPE FORMATION}
\label{sec:Implications}

The observations of the sigmoid and the partial eruption of the active
region in a CME and flare conform very clearly to the expected signatures
of a flux rope with a bald-patch separatrix surface. All key features
theoretically predicted were found in the data. These include
(1) the formation of a continuous S shaped transient sigmoid from a
    diffuse, arcade-like long-lived sigmoid,
(2) the transient sigmoid's triple PIL crossing,
(3) the co-location of the inverse PIL crossing,
    the final merging into a continuous S trace,
    and the area where cancelling strong flux patches converged,
(4) the association with a \emph{partial} filament eruption,
(5) the transient sigmoid's survival through the flare peak, and
(6) the stationary nature of its central part during the eruption.

The combination of features 2, 5, and 6 cannot be attained by field lines
in an erupting sheared arcade. Partial eruptions and the triple PIL
crossing of some field lines within an arcade can be individually
arranged in simulations by applying specific photospheric shear profiles
\cite[e.g.,][]{Antiochos&al1994}. However, a partial eruption that leaves
just the S shaped field lines within the sheared volume in place and
essentially stationary is impossible; the sheared flux would always
participate in the expansion to some degree. Moreover, the MDI data do
not yield any indication of the required shear flow. Consequently, the
findings above are conclusive evidence for the formation of a coronal
flux rope.

The existence of a BPSS in a long-lived, but eventually erupting, sigmoid
was conjectured in another recent case study
\citep{McKenzie&Canfield2008}, based primarily on the J shape of the
sigmoid threads, which by itself is not conclusive, however (see above).
Moreover, only the features 1 and 6 were demonstrated, and no information
was given as to whether the transformation from the double-J to the
single-S appearance of the sigmoid occurred before or in the course of
the eruption.

The formation of the BPSS sigmoid in the active region considered here
began over five hours, and was complete over 80 minutes, prior to the
onset of the flare. A linear backward extrapolation of the CME
height-time data\footnote{\texttt{http://cdaw.gsfc.nasa.gov}}
yields an onset time 10~minutes before flare onset. Since
the CME was rather fast (with a projected velocity of 469~km\,s$^{-1}$ in
spite of its origin near disk centre)
and the associated flare was impulsive, it is very unlikely that the CME
had a gradual initial evolution with a significantly earlier onset time.
Consequently, the data of the sigmoid and
eruption imply that a coronal flux rope existed in the active region at
least one hour, likely over five hours, \emph{prior} to the onset of the
CME.

The data support the gradual arcade-to-rope transformation as the
mechanism of flux rope formation because the evolution of the flux was
dominated by dispersal and cancellation, not by emergence which had
occurred several days earlier.

The transition to the BPSS topology, as seen in soft X-rays, was a
process of at least 14 hours in duration. However, the straight legs of
the J's did not reconnect with each other immediately, and some of their
threads kept a slight separation, albeit decreasing, up to the SXT
exposure at 13:10~UT. The question arises of why a continuous S trace
wasn't visible throughout the transition. We relate this behaviour to the
dominance of the longitudinal flux in the environment of the PIL,
mentioned above, which is characteristic of old flux concentrations
and clearly indicated for the present
active region by the formation of a filament and by the double-J shape of
the long-lived sigmoid. Such flux is nearly unidirectional and can hardly
reconnect inside when perturbed; essentially it represents a barrier to
the approaching flux patches.

Two effects are expected to result in the corona from the observed
localised perturbation. Their relative importance depends on the angle
between the longitudinal and approaching flux, which controls the
strength of reconnection. The approaching flux can slide under the
longitudinal flux (which is rooted remote from the perturbed area),
lifting it gradually, as often observed for filaments prior to an
eruption. The approaching flux will also reconnect in the interface with
the longitudinal flux, illuminating narrow J shaped field line bundles in
the interface. As the approach and reconnection progress, the field lines
lit up lie closer to the PIL, have one footpoint in the approaching flux,
i.e., nearer to the merging point of the sigmoid, and connect to their
remote footpoint in an arc at increasing distance (because they were
brought in from a more remote location and have to encircle a growing
flux rope). These consequences of reconnection are all clearly seen in
the SXT data (Figure~\ref{fig:sigmoid_evolution}). When the approaching
flux eventually reaches direct contact with opposite flux at the PIL
beneath existing longitudinal flux (as indicated by the continued
presence of the filament), the shortened straight legs of the J's merge
into a continuous S by reconnecting with each other near their end
points. The flux they were temporarily rooted in submerges in the
cancellation event, leaving a flux rope with BPSS topology.

\begin{acknowledgements}

We acknowledge the use of data provided by
the Global High Resolution H$\alpha$ Network,
the \textsl{Yohkoh}/SXT, \textsl{SOHO}/MDI, and \textsl{GOES} instruments,
and by the CME catalog held at NASA's CDAW Data Center.
This work was supported by
The Royal Society, STFC, NSF grant ATM 0518218, and NASA grant NNH06AD58I.

\end{acknowledgements}

\end{document}